\documentclass[prb,english,preprint,superscriptaddress,showpacs,preprintnumbers,amsmath,amssymb]{revtex4}
\usepackage[T1]{fontenc}
\usepackage[latin1]{inputenc}
\usepackage{graphicx}

\makeatletter
\usepackage{dcolumn}
\usepackage{bm}
\usepackage{epsfig}

\newcommand{\ignore}[1]{}

\usepackage{babel}
\makeatother
\begin{document}

\title{Electronic Structure of Bilayer Graphene: A Real-space Green's Function Study}

\author{Z.F.Wang}
\affiliation{Hefei National Laboratory for Physical Sciences at
Microscale, University of Science and Technology of China, Hefei,
Anhui 230026, People's Republic of China}

\author{Qunxiang Li}
\thanks{Corresponding author. E-mail: liqun@ustc.edu.cn}
\affiliation{Hefei National Laboratory for Physical Sciences at
Microscale, University of Science and Technology of China, Hefei,
Anhui 230026, People's Republic of China}

\author{Haibin Su}
\affiliation{School of Materials Science and Engineering, Nanyang
Technological University, 50 Nanyang Avenue, 639798, Singapore}

\author{Xiaoping Wang}
\affiliation{Hefei National Laboratory for Physical Sciences at
Microscale, University of Science and Technology of China, Hefei,
Anhui 230026, People's Republic of China}

\author{Q.W.Shi}
\thanks{Corresponding author. E-mail: phsqw@ustc.edu.cn}
\affiliation{Hefei National Laboratory for Physical Sciences at
Microscale, University of Science and Technology of China, Hefei,
Anhui 230026, People's Republic of China}

\author{Jie Chen}
\affiliation{Electrical and Computer Engineering, University of
Alberta, AB T6G 2V4, Canada}

\author{Jinlong Yang}
\affiliation{Hefei National Laboratory for Physical Sciences at
Microscale, University of Science and Technology of China, Hefei,
Anhui 230026, People's Republic of China}

\author{J.G.Hou}
\affiliation{Hefei National Laboratory for Physical Sciences at
Microscale, University of Science and Technology of China, Hefei,
Anhui 230026, People's Republic of China}

\date{\today}

\begin{abstract}
In this paper, a real-space analytical expression for the free
Green's function (propagator) of bilayer graphene is derived based
on the effective-mass approximation. Green's function displays
highly spatial anisotropy with three-fold rotational symmetry. The
calculated local density of states (LDOS) of a perfect bilayer
graphene produces the main features of the observed scanning
tunneling microscopy (STM) images of graphite at low bias voltage.
Some predicted features of the LDOS can be verified by STM
measurements. In addition, we also calculate the LDOS of bilayer
graphene with vacancies by using the multiple-scattering theory
(scatterings are localized around the vacancy of bilayer graphene).
We observe that the interference patterns are determined mainly by
the intrinsic properties of the propagator and the symmetry of the
vacancies.
\end{abstract}

\pacs{73.61Wp, 61.72.Ji, 68.37.Ef}

\maketitle

\section{INTRODUCTION}

Since Novoselov \emph{et al.} fabricated ultra-thin monolayer
graphite devices, \cite{1} the electronic properties of a graphite
monolayer (graphene) have attracted a great deal of research
interest due to the Dirac-type spectrum of charge carriers in its
gap-less semiconductor material. Many interesting properties of
single-layer graphene, such as the Landau quantization, the
defect-induced localization, and spin current states, have been
studied experimentally and theoretically by several research
groups.\cite{2,3,4,5,6,7} These non-orthodox properties, including
massless Dirac fermions around the Dirac points in the first
Brillouin zone, result from graphene's particular band structure.

Recent research attention has focused on the electronic structure of
bilayer and multilayer graphene. \cite{8,9,10,11,12,13,14} Studies
have shown that a bilayer graphene has some unexpected
properties.\cite{11} For example, a bilayer graphene shows anomalies
in its integer quantum Hall effect and in its minimal conductivity
on the order of $e^2/h$.  The common and distinctive electronic
features of single-layer and bilayer graphene are highlighted in
Ref.[15]. Charge carriers in a bilayer graphene are mainly
quasiparticles with a finite density of states at zero energy and
they behave similar to conventional non-relativistic electrons. Like
the relativistic particles or quasiparticles in single-layer
graphene, we can describe these quasiparticles by using spinor
wavefunctions. Although these `massive chiral fermions' do not exist
in the field theory, their existence in condensed-matter physics
offers a unique opportunity to investigate the importance of
chirality and to solve the relativistic tunneling problem.

The unusual physical properties of bilayer graphene are attributed
to two key factors. (i) The relatively weak inter-layer coupling.
Bilayer graphene inherits some properties from single-layer graphene
material, such as the existence of Dirac points in the first
Brillouin zone and the degeneracy of electron and hole bands. (ii)
The special geometry of bilayer graphene with the Bernal stacking
(A-B stacking) between adjacent graphene layers. There are two kinds
of nonequivalent sites (A and B) in each layer as shown in Fig.1(a)
and 1(b). Experimental scanning tunneling microscopy (STM) graphite
images at low bias voltage have verified that only site B is visible
and exhibits a triangular structure. In addition, the orbital
overlap coupling between two adjacent layers is contributed mainly
by the carbon orbitals at site A.\cite{16,17,18} To date, few
analytical studies that attempt to uncover the unique electronic
properties of the bilayer graphene have been done.\cite{12}

In this paper, we first develop an analytical formula  of
electronic structure in a bilayer graphene with the Bernal stacking
based on real-space free Green's function (propagator) and the
effective-mass approximation. We observe that the physical
properties of the bilayer graphene are closely related to its
propagator and the bilayer graphene behaves similar to a massive
chiral Fermions system. Since the local density of states (LDOS) can
be measured by STM, we then compute the LDOS of graphene in various
forms based on  Green's function explicitly. Finally, we highlight
the impact of vacancies' patterns (in terms of their geometry and
symmetry of the computed LDOS) on interferences in STM pictures.

\section{A REAL-SPACE GREEN'S FUNCTION AND ELECTRONIC STRUCTURE OF PERFECT BILAYER GRAPHENE}

In what follows, we derive the analytical expression of the free
Green's function for bilayer graphene in real space based on the
effective-mass approximation. Bilayer graphene consists of two
hexagonal lattice layers coupled by the Bernal stacking as shown in
Fig.1 (a) and (b). In each layer, there are two nonequivalent sites,
A and B. Type-A atoms have direct neighbors in their adjacent layer,
but type-B atoms do not and locate at hollow sites. We assume that
there is one valence electron per carbon atom in a bilayer graphene.
Provided that the difference between wavefunctions orthogonal to the
graphene plane can be neglected, the bilayer graphene can be modeled
as an effective two-dimensional system. Since the physical
properties of graphene are determined by $\pi$ bands near the Dirac
point, only the contribution from the $\pi$ band is considered in
this paper. We further limit our analysis by considering only the
nearest interactions of the graphene's $p_z$ orbitals. The
tight-binding Hamiltonian of the bilayer graphene is
\begin{eqnarray}
H&=&\sum_i\epsilon_{2p}({|i\rangle_{11}}{\langle}i|+{|i\rangle_{22}}{\langle}i|)
\nonumber \\
&
&+V_1\sum_{{\langle}ij\rangle}({|i\rangle_{11}}{\langle}j|+{|i\rangle_{22}}{\langle}j|+h.c.)
\nonumber \\
& &+V_2\sum_{{\langle}ij\rangle}({|i\rangle_{12}}{\langle}j|+h.c.)
\end{eqnarray}

\noindent where ${\langle}\textbf{r}|i\rangle_l$, $l=(1,2)$ is a
wavefunction at site $i$ at layer $l$. $\epsilon_{2p}$ is the
on-site energy, $V_1$ is the nearest hopping parameter within the
layer and $V_2$ is the nearest hopping parameter between two layers.
Our presented calculations are conducted with $\epsilon_{2p}=0 eV$,
$V_1=-3.0 eV$ and $V_2=0.4 eV$.

The Bloch orbits for two nonequivalent sites, A and B, are written
as
${|k_A\rangle_l}=\frac{1}{\sqrt{N}}\sum\limits_{j_A}e^{i\textbf{k}\cdot\textbf{r}_{j_A}}|j_A\rangle_l
\mspace{5mu}$,
${|k_B\rangle}_l=\frac{1}{\sqrt{N}}\sum\limits_{j_B}e^{i\textbf{k}\cdot\textbf{r}_{j_B}}|j_B\rangle_l$.
\noindent The summation covers all sites A and B on layer $l$.
$\textbf{r}_A$ and $\textbf{r}_B$ denote the real-space coordinates
of sites A and B, respectively. Here, N is the number of unit cells
in the crystal. The Hamiltonian can be rewritten as
\begin{eqnarray}
H=\left(\begin{array}{cccc}
\epsilon_{2p} & V_1\mu^* & V_2 & 0\\
V_1\mu & \epsilon_{2p} & 0 & 0\\
V_2 & 0 & \epsilon_{2p} & V_1\mu\\
0 & 0 & V_1\mu^* & \epsilon_{2p}\\
\end{array}\right),
\end{eqnarray}

\noindent where
$\mu=e^{ik_ya}+e^{i(-\frac{\sqrt{3}ak_x}{2}-\frac{ak_y}{2})}+
e^{i(\frac{\sqrt{3}ak_x}{2}-\frac{ak_y}{2})}$ and
$a=1.42${\AA} is the c-c bond length. By defining the retarded
Green's function as, $G_0^{ret}=1/(\varepsilon-H+i\eta)$, where
$\eta$ is an infinitesimal quantity, we obtain the reciprocal space
Green's function for bilayer graphene
\begin{equation}
\begin{split}
\normalsize{G_0^{ret}(\textbf{k},\varepsilon)= \frac{1}{\Delta}
\left(\begin{array}{cccc}
t(t^2-V_1^2\mu\mu^*) & V_1\mu^*(t^2-V_1^2\mu\mu^*) & t^2V_2 & tV_1V_2\mu\\
V_1\mu(t^2-V_1^2\mu\mu^*) & t(t^2-V_1^2\mu\mu^*-V_2^2) & tV_1V_2\mu & V_1^2V_2\mu^2\\
t^2V_2 & tV_1V_2\mu^* & t(t^2-V_1^2\mu\mu^*) & V_1\mu(t^2-V_1^2\mu\mu^*)\\
tV_1V_2\mu^* & V_1^2V_2\mu^{*2} & V_1\mu^*(t^2-V_1^2\mu\mu^*) & t(t^2-V_1^2\mu\mu^*-V_2^2)\\
\end{array}\right)}
\end{split}
\end{equation}

\noindent where
$\Delta=(t^2-V_1^2\mu\mu^*-tV_2)(t^2-V_1^2\mu\mu^*+tV_2)$ and
$t=\varepsilon-\epsilon_{2p}+i\eta$. From $\Delta=0$, the dispersion
relation is expressed as $\varepsilon=\pm \frac {V_2}{2} \pm
\sqrt{(\gamma k)^2+(\frac{V_2}{2})^2}$ with $\gamma=3aV_1/2$, which
is consistent with the previous theoretical result.\cite{12}

The real-space Green's function of the bilayer graphene can be
derived by taking the Fourier transform of
$G_0^{ret}(\textbf{k},\varepsilon)$. For simplicity, these
calculations are carried out around six corners in the first
Brillouin zone within the low-energy region based on the
effective-mass approximation,. Six Dirac points can be divided into
two equivalent sets of points, $K^1,K^3,K^5$ and $K^2,K^4,K^6$,
which are shown in Fig.1(c). They form two $360^\circ$ integrals
around $K^1$ and $K^4$. By using the mathematical relation
$$e^{i\textbf{k}\cdot(\textbf{r}_\mu-\textbf{r}'_\nu)}=J_0(k|\textbf{r}_\mu-\textbf{r}'_\nu|)
+2{\sum_{n=1}^\infty}i^nJ_n(k|\textbf{r}_\mu-\textbf{r}'_\nu|)cos(n\varphi_{r_\mu,r'_\nu}),$$

\noindent where $(\mu,\nu)=(A,B)$, $\varphi_{r_\mu,r'_\nu}$ is the
angle between $\textbf{k}$ and $\textbf{r}_\mu-\textbf{r}'_\nu$, and
$J_n$ is the type-I $n$-order Bessel function. We can simplify the
real-space Green's function of the top layer $(l=1)$ to

\begin{eqnarray}
G_0^{ret}(\textbf{r}_A,\textbf{r}_A',\varepsilon)&=&cos[\textbf{K}^1\cdot(\textbf{r}_A-\textbf{r}'_A)]{\cdot}F_1(|\textbf{r}_A-\textbf{r}'_A|,\varepsilon)\nonumber\\
G_0^{ret}(\textbf{r}_B,\textbf{r}_B',\varepsilon)&=&cos[\textbf{K}^1\cdot(\textbf{r}_B-\textbf{r}'_B)]{\cdot}F_2(|\textbf{r}_B-\textbf{r}'_B|,\varepsilon)\nonumber\\
G_0^{ret}(\textbf{r}_A,\textbf{r}_B',\varepsilon)&=&sin[\textbf{K}^1\cdot(\textbf{r}_A-\textbf{r}'_B)+\alpha_{r_A,r'_B}]\nonumber \\ &&{\cdot}F_3(|\textbf{r}_A-\textbf{r}'_B|,\varepsilon)\nonumber\\
G_0^{ret}(\textbf{r}_B,\textbf{r}_A',\varepsilon)&=&sin[\textbf{K}^1\cdot(\textbf{r}_B-\textbf{r}'_A)-\alpha_{r_B,r'_A}]\nonumber
\\ && {\cdot}F_3(|\textbf{r}_B-\textbf{r}'_A|,\varepsilon),
\end{eqnarray}

\noindent Here $\textbf{K}^1=(4\pi/3\sqrt{3}a,0)$,
$cos\varphi_{r_\mu,r'_\nu}=cos(\theta-\alpha_{r_\mu,r'_\nu})$,
$\theta$ is the angle between the $\textbf{k}$ and x axis.
$\alpha_{r_\mu,r'_\nu}$ is the angle between
$\textbf{r}_\mu-\textbf{r}'_\nu$ and x axis.  The definitions of
$F_1$,$F_2$, and $F_3$ are

\begin{eqnarray}
F_1(|\textbf{r}_\mu-\textbf{r}'_\nu|,\varepsilon)&=&\frac{S}{\pi}
{\int_0^{k_c}}dkkJ_0(k|\textbf{r}_\mu-\textbf{r}'_\nu|) \nonumber \\
&& \times [\frac{t}{t^2-({\gamma}k)^2-tV_2}+\frac{t}{t^2-({\gamma}k)^2+tV_2}],\nonumber\\
F_2(|\textbf{r}_\mu-\textbf{r}'_\nu|,\varepsilon)&=&\frac{S}{\pi}
{\int_0^{k_c}}dkkJ_0(k|\textbf{r}_\mu-\textbf{r}'_\nu|) \nonumber \\
&& \times [\frac{t-V_2}{t^2-({\gamma}k)^2-tV_2}+\frac{t+V_2}{t^2-({\gamma}k)^2+tV_2}], \nonumber\\
F_3(|\textbf{r}_\mu-\textbf{r}'_\nu|,\varepsilon)&=&\frac{S}{\pi}
{\int_0^{k_c}}dkk^2J_1(k|\textbf{r}_\mu-\textbf{r}'_\nu|)
\nonumber\\
&& \times
[\frac{\gamma}{t^2-({\gamma}k)^2-tV_2}+\frac{\gamma}{t^2-({\gamma}k)^2+tV_2}].
\nonumber \\
\end{eqnarray}

\noindent where $S=3\sqrt{3}a^2/2$ is the area of the unit cell in
real space and $k_c$ is the cutoff wave vector. The corresponding
cutoff wave length $2\pi/k_c$ is comparable to the lattice constant
$a$. With the same approach in our previous work to solve electronic
structure in the single-layer graphene, $k_c$ is set to be
$1.71/a$.\cite{19} From Eq.(4), it is clear that the real-space
Green's function of the bilayer graphene is constructed by
multiplying two terms. The first term is spatially anisotropic,
which can be represented by sine and cosine functions determined
from two nonequivalent sets of the Dirac points as shown in
Fig.1(c). The second term including $F_1$, $F_2$ or $F_3$ is
spatially isotropic in real space and depends only on the distance
between two sites.

Many physical properties of bilayer graphene can be deduced from
this explicit expression. Interestingly, when
$k_c\rightarrow\infty$, the functions, $F_1, F_2$ and $F_3$, have
simple analytical forms
\begin{eqnarray}
F_1(|\textbf{r}_\mu-\textbf{r}'_\nu|,\varepsilon)&=&\frac{-S}{\pi\gamma^2}[t
K_0(\frac{it_1}{\gamma}|\textbf{r}_\mu-\textbf{r}'_\nu|)\nonumber
\\ &&+t
K_0(\frac{it_2}{\gamma}|\textbf{r}_\mu-\textbf{r}'_\nu|)], \nonumber\\
F_2(|\textbf{r}_\mu-\textbf{r}'_\nu|,\varepsilon)&=&\frac{-S}{\pi\gamma^2}[(t-V_2)
K_0(\frac{it_1}{\gamma}|\textbf{r}_\mu-\textbf{r}'_\nu|)\nonumber
\\ &&+(t+V_2)
K_0(\frac{it_2}{\gamma}|\textbf{r}_\mu-\textbf{r}'_\nu|)], \nonumber\\
F_3(|\textbf{r}_\mu-\textbf{r}'_\nu|,\varepsilon)&=&\frac{-S}{\pi\gamma^2}[it_1
K_1(\frac{it_1}{\gamma}|\textbf{r}_\mu-\textbf{r}'_\nu|)\nonumber
\\ &&+it_2
K_1(\frac{it_2}{\gamma}|\textbf{r}_\mu-\textbf{r}'_\nu|)],
\end{eqnarray}

\noindent where $t_1=\sqrt{t^2-tV_2}$ and $t_2=\sqrt{t^2+tV_2}$.
$K_0$ and $K_1$ are the zero-th order and the first order terms of
the type-I modified Bessel function, respectively. Compared with our
previous study,\cite{19} the expressions of the Green's function
derived here are the same as those of the single-layer graphene once
the nearest hopping parameter between two layers ($V_2$) is set to
zero. That is to say, it is straightforward to compare theoretical
results of bilayer graphene with previous theoretical and
experimental data of the monolayer graphene by setting
$V_2=0$.\cite{19}

Fig.2 shows the LDOS of bilayer graphene with $\varepsilon$=$-0.1$
eV. Here, the LDOS of site $\textbf{r}_\mu$ is calculated by setting
$\rho_0(\textbf{r}_\mu,\varepsilon)= -\frac{1}{\pi}\mbox{Im}
G_0^{ret}(\textbf{r}_\mu,\textbf{r}_\mu,\varepsilon)$. We
can clearly observe the contrast between the LDOS at site A and that at site B.
Site B is highlighted and forms a triangular lattice with three-fold symmetry.
According to the Tersoff-Hamann model \cite{20} which has been
successfully used to explain experimental results,\cite{21,22} the
STM image can be simulated using the LDOS of the sample surface. It
is reasonable for us to compare the LDOS of bilayer graphene with
previous experimental STM observations of graphite surfaces at low
bias voltage (i.e. 0.3 V). Our calculated LDOS can clearly produce the
main features as shown in several observed STM images reported in Ref.[16-18].

One direct approach to find the difference between bilayer and
monolayer graphene is to study the LDOS at sites A and B for both
monolayer and bilayer graphenes. Fig.3(a) and 3(b) present the LDOS
of one carbon site  (A or B) in the monolayer graphene and two sites
(A and B) in the bilayer graphene, respectively. For a monolayer
graphene, sites A and B are equivalent and thus only one curve of
LDOS is shown in Fig.2(a). However, these two sites are
nonequivalent for a bilayer graphene. Fig.2(b), therefore, clearly
shows that the LDOS of bilayer graphene at site B is greater than
that at site A, when $|\varepsilon|$ is less than approximately 0.4
eV. This LDOS difference leads to the brighter spots at site-B as
shown in Fig.2. When $|\varepsilon|>$ 0.4 eV, the difference between
the LDOS at sites A and B in the bilayer graphene gradually
diminishes. This feature is easy to verify through STM dI/dV mapping
at a relatively large bias voltage. For example, sites A and B can
be both observed in STM dI/dV mapping with the applied bias voltage
of 0.6 V. This phenomenon can also be elaborated by our analytic
expression of Green's function.

The difference between the LDOS of site A and that of site B can be
defined as
\begin{eqnarray}
\Delta\rho_0(\varepsilon)&=&-\frac{1}{\pi}Im[G_0^{ret}(\textbf{r}_B,\textbf{r}_B,\varepsilon)-G_0^{ret}(\textbf{r}_A,\textbf{r}_A,\varepsilon)]
\nonumber
\\ &=&-\frac{1}{\pi}Im[F_2(0,\varepsilon)-F_1(0,\varepsilon)].
\end{eqnarray}

\noindent After some simple math operations, shown in Appendix
Eq.(A-5) and Eq.(A-6), $\Delta\rho_0(\varepsilon)$ can be simplified
to
\begin{eqnarray}
\Delta\rho_0(\varepsilon)= \left\{\begin{array}{cc}
0 & |\varepsilon|>V_2\\
{\frac{SV_2}{2\pi\gamma^2}} & |\varepsilon|<V_2
\end{array}. \right.
\end{eqnarray}

From this equation, the LDOS of the bilayer graphene can be divided
into two regions based on the interlayer hopping parameter $V_2$.
One region is $|\varepsilon|<V_2$, where the LDOS of site B is
larger than that of site A by a constant,
$\frac{SV_2}{2\pi\gamma^2}$. The other region is
$|\varepsilon|>V_2$, where the LDOS of site A equals that of site B.
Our calculations agree well with our numerical results when $V_2$ =
0.4 eV. The calculations show the dispersion relation is
approximately quadratic dependent in the low-energy region and the
results match the previous theoretical results.\cite{12} We find
that the low-energy LDOS of both sites A and B are linearly
proportional to the energy ($\varepsilon$) (refer to the Appendix).
Note that the LDOS of one unit cell at the very low-energy level,
however, is a constant. The LDOS can be expressed as
$\rho(\varepsilon)=[\rho_0(\textbf{r}_A,\textbf{r}_A,\varepsilon)+
\rho_0(\textbf{r}_B,\textbf{r}_B,\varepsilon)]\simeq\frac{SV_2}{2\pi\gamma^2},$
which is the same as those in the conventional two-dimensional
system. This remarkable feature of the bilayer graphene offers a
direct method to measure the inter-layer hopping parameter $V_2$,
which is also the threshold needed to distinguish LDOS contrast in
STM images. The intra-layer hopping parameter $V_1$ can be deduced
based on the LDOS difference measured according to STM images (
$\Delta\rho_0=\frac{SV_2}{2\pi\gamma^2}$ with $\gamma=3aV_1/2$).

\section{LOCAL DENSITY OF STATES OF BILAYER GRAPHENE WITH LATTICE VACANCY}

To validate the accuracy of real-space Green's function derived in
this paper, we use the promoted formula to solve the electronic
structure of a bilayer graphene with a single vacancy. Based on the
tight-binding scheme, a vacancy can be simulated by introducing a
large on-site energy at the vacancy site on a bilayer
graphene.\cite{23} Assuming that the single vacancy locates at site
B with on-site energy $U$, the transport matrix or the T-matrix
\cite{24,25} can be written as $T=U(1-UG_0^{ret})^{-1}$. If U is
large, the T-matrix becomes
\begin{eqnarray}
T(0,0,\varepsilon)&\simeq&-G_0^{ret^{-1}}(0,0,\varepsilon)\nonumber
\\ &=&-F_2^{-1}(0,\varepsilon).
\end{eqnarray}

\noindent Here, the position of the vacancy is set as the coordinate
origin. That is to say, $\textbf{R}_B=0$ in the $x-y$ plane. Using
the Dyson equation, we have
\begin{eqnarray}
G^{ret}(\textbf{r}_\mu,\textbf{r}_\nu,\varepsilon)&=&G_0^{ret}(\textbf{r}_\mu,\textbf{r}_\nu,\varepsilon)
\nonumber
\\ &&+G_0^{ret}(\textbf{r}_\mu,0,\varepsilon)
T(0,0,\varepsilon)G_0^{ret}(0,\textbf{r}_\nu,\varepsilon), \nonumber
\\
\end{eqnarray}

\noindent where $G^{ret}(\textbf{r}_\mu,\textbf{r}_\nu,\varepsilon)$
is the Green's function of bilayer graphene with a single vacancy in
real space. The LDOS on site $\textbf{r}_\mu$ can be determined by
$$\rho(\textbf{r}_\mu,\varepsilon)= -\frac{1}{\pi}\mbox{Im}
G^{ret}(\textbf{r}_\mu,\textbf{r}_\mu,\varepsilon),$$ which can be
simply rewritten as
\begin{eqnarray}
\rho(\textbf{r}_A,\varepsilon)&=&\rho_0(\textbf{r}_A,\varepsilon)+{sin}^2(\textbf{K}^1\cdot\textbf{r}_A+\alpha_{r_A})H_1(\textbf{r}_A,\varepsilon), \nonumber\\
\rho(\textbf{r}_B,\varepsilon)&=&\rho_0(\textbf{r}_B,\varepsilon)+{cos}^2(\textbf{K}^1\cdot\textbf{r}_B)H_2(\textbf{r}_B,\varepsilon).
\label{eq:rho}
\end{eqnarray}

In the above equation,
\begin{eqnarray}
\rho_0(\textbf{r}_A,\varepsilon)&=&-\frac{1}{\pi}Im[F_1(0,\varepsilon)],\nonumber
\\
\rho_0(\textbf{r}_B,\varepsilon)&=&-\frac{1}{\pi}Im[F_2(0,\varepsilon)],
\nonumber \ \\
H_1(\textbf{r}_\mu,\varepsilon)&=&\frac{1}{\pi}Im[\frac{F_3^2(|\textbf{r}_\mu|,\varepsilon)}{F_2(0,\varepsilon)}],
\nonumber
\\
H_2(\textbf{r}_\mu,\varepsilon)&=&\frac{1}{\pi}Im[\frac{F_2^2(|\textbf{r}_\mu|,\varepsilon)}{F_2(0,\varepsilon)}].
\end{eqnarray}

Fig.4(a) shows the calculated LDOS with a single B-site vacancy on
the top sheet of the bilayer graphene. An interesting feature is
clearly observed; bright spots localized near the vacancy around
site A have nice three-fold rotational symmetry. This phenomenon can
be explained based on Eq.(\ref{eq:rho}). Since the function $H_2$ in
the Eq.(\ref{eq:rho}) has a very small value (close to zero), its
magnitude changes little as the distance from the vacancy increases
and thus the magnitude of $\rho(\textbf{r}_B,\varepsilon)$ remains
very small. Although the cosine term in Eq.(\ref{eq:rho}) reaches its
maximum when $r_B=\frac{3\sqrt{3}a}{2}n$, $n=0,\pm1,\pm2\cdots$, the
sine function in $\rho(\textbf{r}_A,\varepsilon)$ reaches its
maximum along directions with angles, $\alpha_{r_A}=30^\circ$,
$90^\circ, 150^\circ$, $210^\circ$, $270^\circ$ and $330^\circ$.
Compared to $H_2$, the value of the function $H_1$ has a relatively
larger value, i.e. 0.4 $eV^{-1}$ at 1.0 {\AA} away from the vacancy
for $\varepsilon$ = -0.1 eV. The function $H_1$, however, decays
rapidly with increasing distance from the vacancy. Note that sites A
along the directions $\alpha_{r_A}=90^\circ$, $210^\circ$ and
$330^\circ$ are located closer to the vacancy than those along the
directions $\alpha_{r_A}=30^\circ$, $150^\circ$ and $270^\circ$. The
calculated site-A LDOS value is between that of the nearest and that
of the next nearest to the B-site vacancy, that is 0.56 eV$^{-1}$
and 0.36 eV$^{-1}$, respectively. Therefore, the sites along the
$\alpha_{r_A}=90^\circ$, $210^\circ$ and $330^\circ$ directions are
brighter than those along the $\alpha_{r_A}=30^\circ$, $150^\circ$
and $270^\circ$ directions. These bright spots show the localized
character of the region surrounding the vacancy. By comparing to the
dI/dV image of a vacancy on the single-layer graphene \cite{19}, the
enhanced localization of LDOS in the bilayer graphene is caused by
the additional inter-layer channel propagated from the vacancy
point. Green's function can oscillate in a significantly longer
distance starting from the vacancy in the monolayer graphene than in
the bilayer graphene. The asymptotic behavior of $H_1$ can help us
understand this phenomenon. When the distance is longer than 4.0
{\AA}, $H_1$ is less than 0.1 eV$^{-1}$ and thus these bright spots
show the localized character of the region within nearest lattice
surrounding the vacancy.

The LDOS with  vacancy near site-A is also simulated as shown in
Fig. 4(b). We observe the similar features that the vacancy around
site-B has bright spots with three-fold symmetry. The calculated
site-B LDOS value is between that of the nearest and that of the
next nearest to the site-A vacancy, which is between 0.35 eV$^{-1}$
and 0.23 eV$^{-1}$.

The extension of our calculation to consider several vacancies on
bilayer graphene is straightforward. The scattering T-matrix in
Eq.(10) includes all contributions from vacancies. Similar to a
single-vacancy case, a large on-site energy $U$ for all vacancy
sites is used in our simulations. Fig.4(c) shows the LDOS near an
AB-type vacancy (a couple of the neighboring sites A and B) on a
bilayer graphene surface. The LDOS is still localized around the
vacancies. However, the clear three-fold rotational symmetry
disappears in this case. The bright spots localized above the
vacancy correspond to the B sites. Note that the LDOS at the bright
spots is about 0.03 eV$^{-1}$, which is less than those in Fig.4(a)
and (b) by an order of magnitude. This phenomenon results in the
destructive interference due to the multiple-scattering caused by
two vacancies. The asymmetric pattern in Fig. 4(c) reflects the
residual contribution from the destructive interference since two sites A and
B are nonequivalent in the bilayer graphene. Our result suggests
that the interference pattern is determined mainly by vacancies,
provided that the spatial anisotropy of the Green's function is
fixed. The symmetry will be lost in LDOS whenever vacancies break
the three-fold rotational symmetry in bilayer graphene.

\section{CONCLUSION}
In this paper, an analytical form of the real-space Green's function
(propagator) of bilayer graphene is constructed based on the
effective-mass approximation. Green's function demonstrates an
elegant spatial anisotropy with a three-fold symmetry for
defect-free bilayer graphene. The LDOS of the bilayer graphene
determines the main features of experimental STM images on graphite
surfaces with low-bias voltage. The predicted features according to
our simulated results can be verified by STM measurements. For
example, two nonequivalent atomic sites can be observed in STM dI/dV
images with different bias voltages. The information of interlayer
and intralayer hopping strength can be deduced based on the contrast
of STM images. Moreover, We also calculate the LDOS of bilayer
graphene with vacancies by using the multiple-scattering theory. The
interference patterns are determined mainly by the properties of
Green's function and the symmetry of the vacancies. Once the
vacancies break the intrinsic symmetry of the graphene, the
three-fold rotational symmetry of the LDOS vanishes. Our model
provides exact results near the Dirac points. We have discovered
some interesting STM patterns of the second layer, but these results
will be published elsewhere. In this paper, the bilayer graphene is
described by using the simple non-interactive tight-binding scheme.
We are currently investigating how Coulomb interaction impacts the
electronic structure of the bilayer graphene.

\section*{ACKNOWLEDGMENTS}
This work is partially supported by the National Natural Science
Foundation of China under Grants 10274076, 10674121, 20533030,
20303015 and 50121202, by National Key Basic Research Program under
Grant No. 2006CB0L1200, by the USTC-HP HPC project, and by the SCCAS
and Shanghai Supercomputer Center. Work at NTU is supported in part
by COE-SUG grant (No. M58070001). Jie Chen would like to acknowledge
the funding support from the Discovery program of Natural Sciences
and Engineering Research Council of Canada (No. 245680).

\renewcommand{\theequation}{A-\arabic{equation}}
\setcounter{equation}{0}  
\section*{APPENDIX}  

From Eq.(3), Green's functions at the top layer in reciprocal
space are expressed as
\begin{eqnarray}
G_{0AA}^{ret}(\textbf{k},\varepsilon)=\frac{1}{2}[\frac{t}{t^2-V_1^2\mu\mu^*-tV_2}+\frac{t}{t^2-V_1^2\mu\mu^*+tV_2}]\nonumber\\
G_{0BB}^{ret}(\textbf{k},\varepsilon)=\frac{1}{2}[\frac{t-V_2}{t^2-V_1^2\mu\mu^*-tV_2}+\frac{t+V_2}{t^2-V_1^2\mu\mu^*+tV_2}]\nonumber\\
G_{0AB}^{ret}(\textbf{k},\varepsilon)=\frac{1}{2}[\frac{V_1\mu^*}{t^2-V_1^2\mu\mu^*-tV_2}+\frac{V_1\mu^*}{t^2-V_1^2\mu\mu^*+tV_2}]\nonumber\\
G_{0BA}^{ret}(\textbf{k},\varepsilon)=\frac{1}{2}[\frac{V_1\mu}{t^2-V_1^2\mu\mu^*-tV_2}+\frac{V_1\mu}{t^2-V_1^2\mu\mu^*+tV_2}]
\nonumber \\
\end{eqnarray}

\noindent By taking the Fourier transform of
$G_{0\mu\nu}^{ret}(\textbf{k},\varepsilon)$ in the first Brillouin
zone (1BZ), we can obtain the exact expression of  Green's
function in real space for bilayer graphene
$G_0^{ret}(\textbf{r}_\mu,\textbf{r}'_\nu,\varepsilon)=\int_{1BZ}d\textbf{k}G_{0\mu\nu}^{ret}(\textbf{k},\varepsilon)e^{i\textbf{k}\cdot(\textbf{r}_\mu-\textbf{r}'_\nu)}$,
where $\mu$ and $\nu$ are for site-A or site-B atoms, respectively. Based on the
effective-mass approximation, we can sum \emph{k} points near the
six corners (labeled by $i=1\sim6$) in the first Brillouin zone. The
real-space bilayer graphene Green's function can be written as
\begin{eqnarray}
G_0^{ret}(\textbf{r}_A,\textbf{r}'_A,\varepsilon)&=&{\frac{S}{(2\pi)^2}}\sum\limits_{i=1}^6\int
dk_x^idk_y^i
[\frac{t}{t^2-V_1^2\mu_i\mu_i^*-tV_2}+\frac{t}{t^2-V_1^2\mu_i\mu_i^*+tV_2}]e^{i(\textbf{K}^i+\textbf{k}^i)\cdot(\textbf{r}_A-\textbf{r}'_A)}\nonumber\\
G_0^{ret}(\textbf{r}_B,\textbf{r}'_B,\varepsilon)&=&{\frac{S}{(2\pi)^2}}\sum\limits_{i=1}^6\int
dk_x^idk_y^i
[\frac{t-V_2}{t^2-V_1^2\mu_i\mu_i^*-tV_2}+\frac{t+V_2}{t^2-V_1^2\mu_i\mu_i^*+tV_2}]e^{i(\textbf{K}^i+\textbf{k}^i)\cdot(\textbf{r}_B-\textbf{r}'_B)}\nonumber\\
G_0^{ret}(\textbf{r}_A,\textbf{r}'_B,\varepsilon)&=&{\frac{S}{(2\pi)^2}}\sum\limits_{i=1}^6\int
dk_x^idk_y^i
[\frac{V_1\mu_i^*}{t^2-V_1^2\mu_i\mu_i^*-tV_2}+\frac{V_1\mu_i^*}{t^2-V_1^2\mu_i\mu_i^*+tV_2}]e^{i(\textbf{K}^i+\textbf{k}^i)\cdot(\textbf{r}_A-\textbf{r}'_B)}\nonumber\\
G_0^{ret}(\textbf{r}_B,\textbf{r}'_A,\varepsilon)&=&{\frac{S}{(2\pi)^2}}\sum\limits_{i=1}^6\int
dk_x^idk_y^i
[\frac{V_1\mu_i}{t^2-V_1^2\mu_i\mu_i^*-tV_2}+\frac{V_1\mu_i}{t^2-V_1^2\mu_i\mu_i^*+tV_2}]e^{i(\textbf{K}^i+\textbf{k}^i)\cdot(\textbf{r}_B-\textbf{r}'_A)}
\nonumber \\
\end{eqnarray}

In Eq.(A-2), the integral around $K^1,K^3,K^5$ and $K^2,K^4,K^6$ can
be summed together separately to form two $360^\circ$ integrals
around $K^1$ and $K^4$, that is

\begin{eqnarray}
G_0^{ret}(\textbf{r}_A,\textbf{r}'_A,\varepsilon)&=&{\frac{S}{(2\pi)^2}}[e^{i\textbf{K}^1(\textbf{r}_A-\textbf{r}'_A)}+e^{i\textbf{K}^4(\textbf{r}_A-\textbf{r}'_A)}]
\nonumber\\ &&\times \int_0^{2\pi}d\theta\int_0^{k_c}dkk
{[\frac{t}{t^2-(\gamma k)^2-tV_2}+\frac{t}{t^2-(\gamma k)^2+tV_2}]}e^{i\textbf{k}\cdot(\textbf{r}_A-\textbf{r}'_A)}\nonumber\\
G_0^{ret}(\textbf{r}_B,\textbf{r}'_B,\varepsilon)&=&{\frac{S}{(2\pi)^2}}[e^{i\textbf{K}^1(\textbf{r}_B-\textbf{r}'_B)}+e^{i\textbf{K}^4(\textbf{r}_B-\textbf{r}'_B)}]
\nonumber\\ &&\times \int_0^{2\pi}d\theta\int_0^{k_c}dkk
{[\frac{t-V_2}{t^2-(\gamma k)^2-tV_2}+\frac{t+V_2}{t^2-(\gamma k)^2+tV_2}]}e^{i\textbf{k}\cdot(\textbf{r}_B-\textbf{r}'_B)}\nonumber\\
G_0^{ret}(\textbf{r}_A,\textbf{r}'_B,\varepsilon)&=&{\frac{S}{(2\pi)^2}}\{
e^{i\textbf{K}^1(\textbf{r}_A-\textbf{r}'_B)} \nonumber\\ &&\times
\int_0^{2\pi}d\theta\int_0^{k_c}dkk^2\gamma(-isin\theta-cos\theta)
{[\frac{1}{t^2-(\gamma k)^2-tV_2}+\frac{1}{t^2-(\gamma k)^2+tV_2}]}e^{i\textbf{k}\cdot(\textbf{r}_A-\textbf{r}'_B)}\nonumber\\
&&+e^{i\textbf{K}^4(\textbf{r}_A-\textbf{r}'_B)} \nonumber\\
&&\times \int_0^{2\pi}d\theta
\int_0^{k_c}dkk^2\gamma(-isin\theta+cos\theta)
{[\frac{1}{t^2-(\gamma k)^2-tV_2}+\frac{1}{t^2-(\gamma k)^2+tV_2}]}e^{i\textbf{k}\cdot(\textbf{r}_A-\textbf{r}'_B)}\}\nonumber\\
G_0^{ret}(\textbf{r}_B,\textbf{r}'_A,\varepsilon)&=&{\frac{S}{(2\pi)^2}}\{
e^{i\textbf{K}^1(\textbf{r}_B-\textbf{r}'_A)}\nonumber\\&& \times
\int_0^{2\pi}d\theta\int_0^{k_c}dkk^2\gamma(isin\theta-cos\theta)
{[\frac{1}{t^2-(\gamma k)^2-tV_2}+\frac{1}{t^2-(\gamma k)^2+tV_2}]}e^{i\textbf{k}\cdot(\textbf{r}_B-\textbf{r}'_A)}\nonumber\\
&&+e^{i\textbf{K}^4(\textbf{r}_B-\textbf{r}'_A)}\int_0^{2\pi}d\theta
\nonumber\\ &&\times \int_0^{k_c}dkk^2\gamma(isin\theta+cos\theta)
{[\frac{1}{t^2-(\gamma k)^2-tV_2}+\frac{1}{t^2-(\gamma
k)^2+tV_2}]}e^{i\textbf{k}\cdot(\textbf{r}_B-\textbf{r}'_A)}\} \nonumber\\
\end{eqnarray}

\noindent By using relation
$$e^{i\textbf{k}\cdot(\textbf{r}_\mu-\textbf{r}'_\nu)}=J_0(k|\textbf{r}_\mu-\textbf{r}'_\nu|)
+2{\sum_{n=1}^\infty}i^nJ_n(k|\textbf{r}_\mu-\textbf{r}'_\nu|)cos(n\varphi_{r_\mu,r'_\nu}),$$
and integrating the angle part of Eq.(A-3), we can easily get Eqs.(4) and (5).
If we let $k_c\rightarrow\infty$, Eq.(5) is
simplified to Eq.(6).

Next, we briefly derive Eq.(8). At the same site A or
B, the corresponding Green's functions are:

\begin{eqnarray}
G_0^{ret}(\textbf{r}_A,\textbf{r}_A,\varepsilon)=\frac{S}{\pi}
{\int_0^{k_c}}dkk[\frac{t}{t^2-({\gamma}k)^2-tV_2}+\frac{t}{t^2-({\gamma}k)^2+tV_2}],\nonumber\\
G_0^{ret}(\textbf{r}_B,\textbf{r}_B,\varepsilon)=\frac{S}{\pi}
{\int_0^{k_c}}dkk[\frac{t-V_2}{t^2-({\gamma}k)^2-tV_2}+\frac{t+V_2}{t^2-({\gamma}k)^2+tV_2}].\nonumber\\
\end{eqnarray}

\noindent The LDOS at sites A and B are
\begin{eqnarray}
\rho_0(\textbf{r}_A,\textbf{r}_A,\varepsilon)&=&-\frac{1}{\pi}Im[G_0^{ret}(\textbf{r}_A,\textbf{r}_A,\varepsilon)]\nonumber\\
&=&-\frac{S}{2(\pi\gamma)^2}Im\int_0^{k_c}d[\sqrt{(\gamma k)^2+(\frac{V_2}{2})^2}]\nonumber\\
&&\{\frac{t}{\varepsilon+i\eta-\frac{V_2}{2}-\sqrt{(\gamma
k)^2+(\frac{V_2}{2})^2}}-
\frac{t}{\varepsilon+i\eta-\frac{V_2}{2}+\sqrt{(\gamma k)^2+(\frac{V_2}{2})^2}}+\nonumber\\
&&\frac{t}{\varepsilon+i\eta+\frac{V_2}{2}-\sqrt{(\gamma
k)^2+(\frac{V_2}{2})^2}}-
\frac{t}{\varepsilon+i\eta+\frac{V_2}{2}+\sqrt{(\gamma k)^2+(\frac{V_2}{2})^2}}\}\nonumber\\
&=&\left\{\begin{array}{cc}
{\frac{S|\varepsilon|}{\pi\gamma^2}} & |\varepsilon|>V_2\\
{\frac{S|\varepsilon|}{2\pi\gamma^2}} &  |\varepsilon|<V_2\\
\end{array}. \right.
\end{eqnarray}

\begin{eqnarray}
\rho_0(\textbf{r}_B,\textbf{r}_B,\varepsilon)&=&-\frac{1}{\pi}Im[G_0^{ret}(\textbf{r}_B,\textbf{r}_B,\varepsilon)]\nonumber\\
&=&-\frac{S}{2(\pi\gamma)^2}Im\int_0^{k_c}d[\sqrt{(\gamma k)^2+(\frac{V_2}{2})^2}]\nonumber\\
&&\{\frac{t-V_2}{\varepsilon+i\eta-\frac{V_2}{2}-\sqrt{(\gamma
k)^2+(\frac{V_2}{2})^2}}-
\frac{t-V_2}{\varepsilon+i\eta-\frac{V_2}{2}+\sqrt{(\gamma k)^2+(\frac{V_2}{2})^2}}+\nonumber\\
&&\frac{t+V_2}{\varepsilon+i\eta+\frac{V_2}{2}-\sqrt{(\gamma
k)^2+(\frac{V_2}{2})^2}}-
\frac{t+V_2}{\varepsilon+i\eta+\frac{V_2}{2}+\sqrt{(\gamma k)^2+(\frac{V_2}{2})^2}}\}\nonumber\\
&=&\left\{\begin{array}{cc}
{\frac{S|\varepsilon|}{\pi\gamma^2}} & |\varepsilon|>V_2\\
{\frac{S(|\varepsilon|+V_2)}{2\pi\gamma^2}} &  |\varepsilon|<V_2\\
\end{array}. \right.
\end{eqnarray}

\noindent Then, the difference between LDOS at A and B sites is
\begin{eqnarray}
\Delta\rho_0(\varepsilon)&=&\rho_0(\textbf{r}_B,\textbf{r}_B,\varepsilon)-\rho_0(\textbf{r}_A,\textbf{r}_A,\varepsilon)\nonumber\\
&=&\left\{\begin{array}{cc}
0 & |\varepsilon|>V_2\\
{\frac{SV_2}{2\pi\gamma^2}} & |\varepsilon|<V_2
\end{array}. \right.
\end{eqnarray}

\newpage{} %
\begin{figure}[htbp]
 \includegraphics[width=7.5cm]{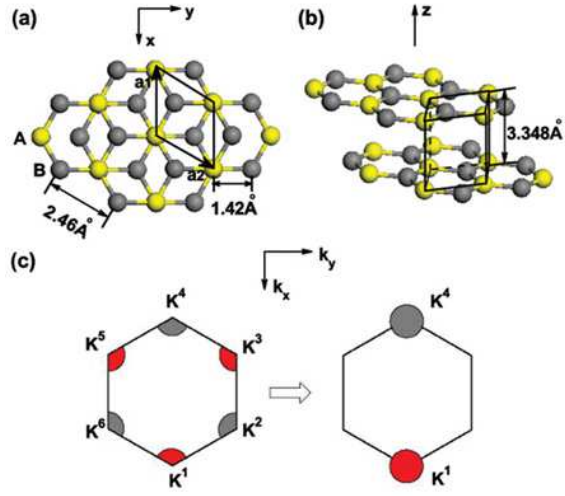}
\caption{(color online) The crystal structure of a bilayer graphene.
The unit cell consists of two layers with two nonequivalent sites: A
(yellow) and B (gray). (a) Top view with surface unit vectors
\textit{a1} and \textit{a2}, (b) Side view. (c) The first Brillouin
zone of a bilayer graphene, where $K^{1}{\sim}K^{6}$ are the Dirac
points. These Dirac points can be further divided into two
nonequivalent sets, $K^1\sim(K^{1},K^{3},K^{5})$ and
$K^4\sim(K^{2},K^{4},K^{6})$. }
\end{figure}

\newpage{} %
\begin{figure}[htbp]
\includegraphics[width=7.5cm]{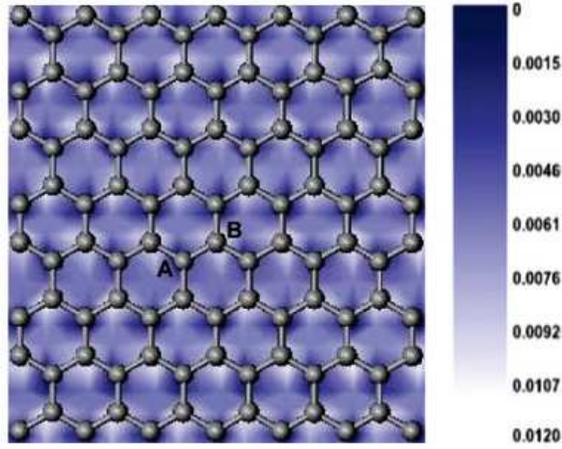}
\caption{(color online) LDOS of bilayer graphene with
$\varepsilon$=-0.1 eV. The LDOS at B site is larger than that at A
site, represented by the brighter contour.}
\end{figure}

\newpage{} %
\begin{figure}[htbp]
\includegraphics[width=7.5cm]{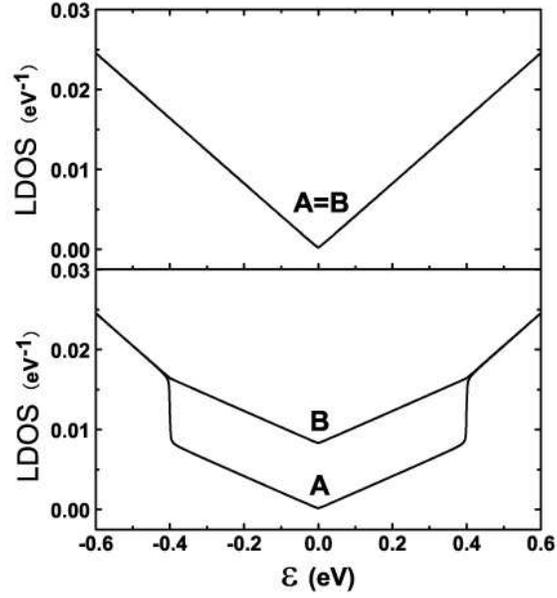}
\caption{(a) LDOS of monolayer graphene on one site (sites A and B
are equivalent). (b) LDOS of bilayer graphene on two sites (sites A
and B are nonequivalent).}
\end{figure}

\newpage{} %
\begin{figure}[htbp]
\includegraphics[width=7.5cm]{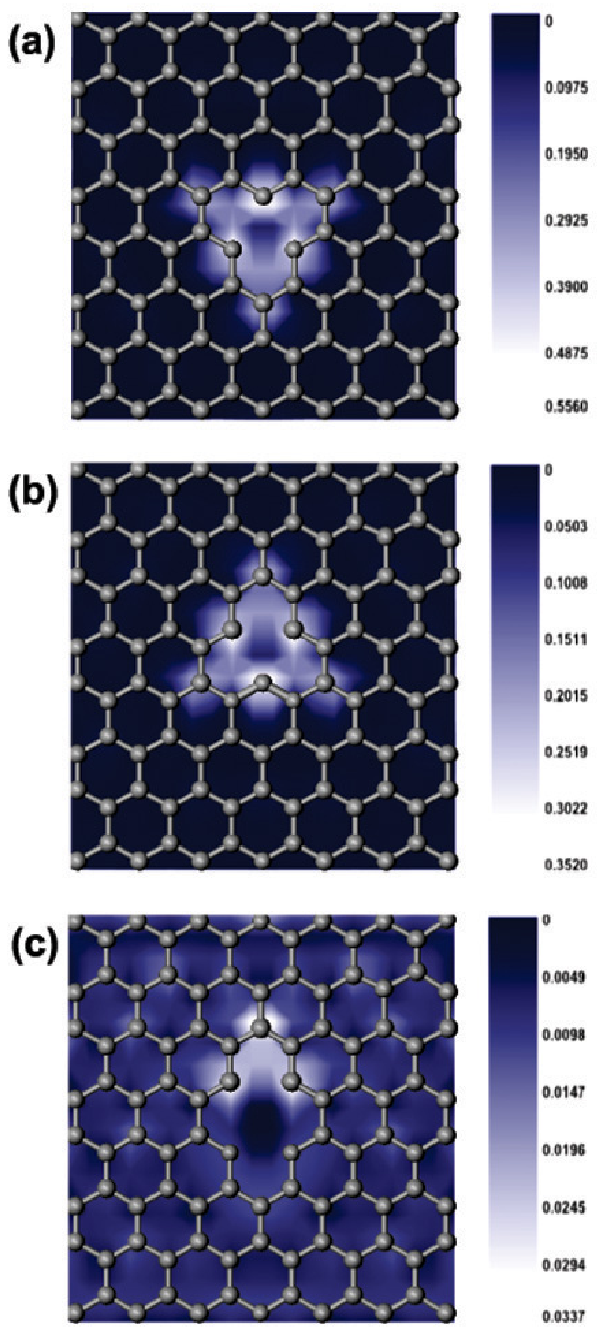}
\caption{(color online) LDOS of vacancies in bilayer graphene. (a)
single B-site vacancy; (b) A-site vacancy; (c) a pair of AB vacancy.
Here,$\varepsilon$=-0.1 eV}
\end{figure}


\begin{references}
\bibitem{1}
K. S. Novoselov, A. K. Geim, S. V. Morozov, D. Jiang, M. I.
Katsnelson, I. V. Grigorieva, S. V. Dubonos, and A. A. Firsov,
Nature \textbf{438},197 (2005).

\bibitem{2}
T. Matsui, H. Kambara, Y. Niimi, K. Tagami, M. Tsukada, and Hiroshi
Fukuyama, Phys. Rev. Lett. \textbf{94},226403 (2005).

\bibitem{3}
Vitor M. Pereira, F. Guinea, J. M. B. Lopes dos Santos, N. M. R.
Peres, and A. H. Castro Neto, Phys. Rev. Lett. \textbf{96},036801
(2006).

\bibitem{4}
J. Tworzyd{\l}o, B. Trauzettel, M. Titov, A. Rycerz, and C.W. J.
Beenakker, Phys. Rev. Lett. \textbf{96},246802 (2006).

\bibitem{5}
N. M. R. Peres, F. Guinea, and A. H. Castro Neto, Phys. Rev. B
\textbf{73},125411 (2006).

\bibitem{6}
Y. Zhang, Z. Jiang, J. P. Small, M. S. Purewal, Y.-W. Tan, M.
Fazlollahi, J. D. Chudow, J. A. Jaszczak, H. L. Stormer, and P. Kim,
Phys. Rev. Lett. \textbf{96},136806 (2006).

\bibitem{7}
Dmitry A. Abanin, Patrick A. Lee, and Leonid S. Levitov, Phys. Rev.
Lett. \textbf{96},176803 (2006).

\bibitem{8}
Claire Berger, Zhimin Song, Tianbo Li, Xuebin Li, Asmerom Y.
Ogbazghi, Rui Feng, Zhenting Dai, Alexei N. Marchenkov, Edward H.
Conrad, Phillip N. First, and Walt A. de Heer, J. Phys. Chem. B
\textbf{108}, 19912 (2004).

\bibitem{9}
Claire Berger, Zhimin Song, Xuebin Li, Xiaosong Wu, Nate Brown,
C\'{e}cile Naud, Didier Mayou, Tianbo Li, Joanna Hass, Alexei N.
Marchenkov, Edward H. Conrad, Phillip N. First, and Walt A. de Heer,
Science \textbf{312}, 1191 (2006)

\bibitem{10}
Johan Nilsson, A. H. Castro Neto, F. Guinea, N. M. R. Peres,
cond-mat/0604106.

\bibitem{11}
K. S. Novoselov, E. McCacn, S. V. Morozov, V. I. Fal'ko, M. I.
Katsnelson, U. Zeitler, D. Jiang, F. Schedin, A. K. Geim, Nature
Phys. \textbf{2},177 (2006).

\bibitem{12}
Edward McCann and Vladimir I. Fal'ko, Phys. Rev. Lett.
\textbf{96},086805 (2006).

\bibitem{13}
Mikito Koshino and Tsuneya Ando, Phys. Rev. B \textbf{73},245403
(2006).

\bibitem{14}
Johan Nilsson, A. H. Castro Neto, F. Guinea, N. M. R. Peres,
cond-mat/0607343.

\bibitem{15}
M. I. Katsnelson, K. S. Novoselov, A. K. Geim, Nature Phys.
\textbf{2}, 620 (2006).

\bibitem{16}
David Tom\'{a}nek, Steven G. Louie, H. Jonathon Mamin, David W.
Abraham, Ruth Ellen Thomson, Eric Ganz, and John Clarke, Phys. Rev.
B \textbf{35},7790 (1987).

\bibitem{17}
Stefan Hembacher, Franz J. Giessibl, Jochen Mannhart, and Calvin F.
Quate, Proc. Natl. Acad. Sci. U.S.A. \textbf{100}, 12539 (2003).

\bibitem{18}
Yongfeng Wang, Yingchun Ye, Kai Wu, Surf. Sci. \textbf{600},729
(2006).

\bibitem{19}
Z. F. Wang, Ruoxi Xiang, Q. W. Shi, Jinlong Yang, Xiaoping Wang, J.
G. Hou, and Jie Chen, Phys. Rev. B \textbf{74}, 125417 (2006).

\bibitem{20}
J. Tersoff and D. R. Hamann, Phys. Rev. B \textbf{31}, 805 (1985).

\bibitem{21}
J. G. Hou, Jinlong Yang, Haiqian Wang, Qunxiang Li, Changgan Zeng,
Hai Lin, Bing Wang, D. M. Chen, and Qingshi Zhu, Phys. Rev. Lett.
\textbf{83}, 3001 (1999); J.G.Hou, Jinlong Yang, Haiqian Wang,
Qunxiang Li, Changgan Zeng, Lanfeng Yuan, Bing Wang, D. M. Chen,
Qingshi Zhu, Nature \textbf{409}, 304 (2001); Aidi Zhao, Qunxiang
Li, Lan Chen, Hongjun Xiang, Weihua Wang, Shuan Pan, Bing Wang,
Jinlong Yang, J. G. Hou, and Qingshi Zhu, Science \textbf{309} 1542
(2005).

\bibitem{22} Qunxiang Li and Xiao Hu, Phys. Rev. B
\textbf{74}, 035414 (2006).

\bibitem{23}
Tsuneya Ando, Takeshi Nakanishi and Masatsura Igami, J. Phys. Soc.
Jpn. \textbf{68}, 3994 (1999).

\bibitem{24}
Qiang-Hua Wang and Dung-Hai Lee, Phys. Rev. B \textbf{67},020511(R)
(2003).

\bibitem{25}
J. M. Byers, M. E. Flatt\'{e}, and D. J. Scalapino, Phys. Rev. Lett.
\textbf{71},3363 (1993).

\end{references}
\end{document}